\documentclass[aps,prd,twocolumn,tenpt,eqsecnum,%
              nofootinbib]{revtex4}
\usepackage{graphicx,epsf,amssymb}
\newcommand{\be}{\begin{equation}}
\newcommand{\ee}{\end{equation}}

\newcommand{\bea}{\begin{eqnarray}}
\newcommand{\eea}{\end{eqnarray}}

\newif\ifdraft
\drafttrue
\newif\ifpreprint
\preprinttrue

\def\eqn#1{eq.~(\ref{#1})}

\def\fig#1{fig.~{\ref{#1}}}

\def\Fig#1{Figure~{\ref{#1}}}

\def\del{\partial}

\newbox\charbox
\newbox\slabox
\def\s#1{{      % Feynman slash
        \setbox\charbox=\hbox{$#1$}
        \setbox\slabox=\hbox{$/$}
        \dimen\charbox=\ht\slabox
        \advance\dimen\charbox by -\dp\slabox
        \advance\dimen\charbox by -\ht\charbox
        \advance\dimen\charbox by \dp\charbox
        \divide\dimen\charbox by 2
        \raise-\dimen\charbox\hbox to \wd\charbox{\hss/\hss}
        \llap{$#1$}
}}

\def\qb{{\overline {\kern-0.7pt q\kern -0.7pt}}}

\def\e{\epsilon}

\def\Ord{{\cal O}}

\newbox\ourfigbox
\def\SizedFigureWithCaption#1#2#3{%
\setbox\ourfigbox
  \hbox{\hss\epsfxsize #1 \epsfbox{#2}\hss}
\hbox to \wd\ourfigbox{\vbox{\noindent\copy\ourfigbox\break
\vskip -6mm      \hbox to \wd\ourfigbox{\hss#3\hss}}}
}

\begin{document}
\hfuzz 15 pt

%%%%%%%%%%%%%%%%%%%%%%%%%%%%%%%

\ifpreprint
%\noindent ~~~\hfill SLAC--PUB--13054 
\leftline{SLAC--PUB--13054}
\fi

\title{Hard QCD Processes at Colliders}

\author{Lance J. Dixon\thanks{Research supported by 
the US Department of Energy under contract DE--AC02--76SF00515}}
\affiliation{Stanford Linear Accelerator Center \\ 
              Stanford University \\
             Stanford, CA 94309, USA}

\begin{abstract}
Recent developments in the study of hard QCD processes at colliders
are reviewed, in the context of the imminent startup of the LHC.
\vskip0.3in
\centerline{\it Invited talk presented at the XXIII International Symposium} 
\centerline{\it on Lepton and Photon Interactions at High Energy (LP07)}
\centerline{\it August 13 - 18, 2007, Daegu, Korea}
\end{abstract}

\maketitle

\renewcommand{\thefootnote}{\arabic{footnote}}
\setcounter{footnote}{0}

%%%%%%%%%%% Section I  %%%%%%%%%%%%%%%%%%%%%%%%%%%%%%%%%%%%%%

\section{Introduction}
\label{IntroSection}

Next year, the Large Hadron Collider
(LHC) will begin operation at CERN, 
colliding protons at a center-of-mass energy of 14~TeV, 
seven times greater than that currently available 
in $p\bar{p}$ collisions at Fermilab's Tevatron.
The LHC luminosity should also be a factor of 10
to 100 greater than the Tevatron's.  The combined rise 
in energy and luminosity represents the opening of a new window
into electroweak-scale physics.   There will be copious production
of heavy states such as electroweak vector bosons and top quarks, 
and, it is anticipated, Higgs boson(s) and physics beyond the
Standard Model.

Are we ready to exploit this new window?
That is, is Standard Model physics at the LHC understood well enough to 
confidently extract physics beyond the Standard Model?
Of course the physics at the LHC {\it is} largely the physics
of hard QCD processes (with important contributions from soft regimes
as well).  QCD itself will be probed in unprecedented regimes
and with unprecedented statistics.  QCD governs the production
of electroweak states in and beyond the Standard Model.
Pure jet final states also need to be understood, because jet production
rates are so large.  Jets can fake electroweak signatures
such as leptons or photons, and can fake missing energy via mismeasurement.
In general, QCD (plus electroweak) hard processes form significant 
backgrounds to almost all new physics searches at the LHC, 
as well as at the Tevatron.

To be sure that our understanding of hard QCD at the LHC is good enough,
it is important to check QCD predictions against current
data from the Tevatron and HERA.  The main experimental inputs
are the strong coupling, $\alpha_s(M_Z)$, and the parton distribution
functions.  

Recent progress in computing three-jet observables
in $e^+e^-$ annihilation at next-to-next-to-leading order 
(NNLO)~\cite{GDGGHThrust}, to be described below, promises
to improve the experimental uncertainty in $\alpha_s$
as determined from $e^+e^-$ event shapes.  
On the other hand, the uncertainty
in the current world average~\cite{Bethke04},
\be 
\alpha_s(M_Z) = 0.1182 \pm 0.0027\,,
\label{alphasworld04}
\ee
is already small, in comparison with other uncertainties, 
for almost all LHC processes.
Parton distribution functions (pdfs), and their uncertainties, are critical
to all QCD predictions, but as they were reviewed in the talks
by Gwenlan~\cite{GwenlanLP07} and Diehl~\cite{DiehlLP07},
I will not dwell on them here.  Here I will focus
more on our theoretical understanding of the hard, short-distance
structure of QCD processes, assessed when possible against
Tevatron and HERA data for various processes and regimes.

Because hard QCD is a vast subject, this talk will only scratch
the surface.  I will say next to nothing about several important 
subjects covered in part by other speakers, 
such as the high-energy (BFKL) limit, small-$x$ physics and parton
saturation~\cite{DiehlLP07}; 
(hard) diffraction~\cite{DiehlLP07,RostovtsevLP07};
and heavy quark production~\cite{GwenlanLP07,ErbacherLP07}.
I will also not be able to cover multiple parton scattering
and the underlying event; the substantial recent progress in
matching leading-order QCD predictions with parton showers;
other Monte Carlo developments; insights for collider physics
being developed via soft collinear effective theory; and 
new techniques for computing one-loop QCD amplitudes with
many external legs.

Section II of this article outlines the framework and basic elements of 
fixed-order QCD computations and briefly discusses resummation.
Section III describes a state-of-the-art
application to $e^+e^-$ annihilation, the thrust distribution
at NNLO~\cite{GDGGHThrust}.  In Section IV, developments in 
the theory of Higgs production at hadron colliders are summarized, 
along with selected decay channels and some backgrounds thereof.
Section V is devoted to jet physics:  definitions, substructure,
rates and distributions.  Section VI covers the production of a
vector boson in association with jets.  Section VII discusses
the production of a top quark pair plus a jet, and the $t\bar{t}$
forward-backward asymmetry.  In Section VIII I conclude.

%%%%%%%%%%% Section II  %%%%%%%%%%%%%%%%%%%%%%%%%%%%%%%%%%%%%%

\section{Preliminaries and theoretical tools}
\label{PrelimSection}

Asymptotic freedom in QCD~\cite{GWP} guarantees that at short distances
(large transverse momenta) the partons in the proton are almost free,
and are sampled essentially one at a time in hard collisions.
This picture leads to the QCD-improved parton model, in which 
the hadronic cross section for production of a final state $X$
factorizes into products of pdfs $f_a$ and partonic cross sections 
$\hat\sigma^{ab\to X}$,
\bea
&&\hskip-1.0cm
\sigma^{pp\to X}(s;\alpha_s,\mu_F,\mu_R) 
\nonumber\\
&=& \sum_{a,b} \int_0^1 dx_1 dx_2\  
f_a(x_1,\alpha_s,\mu_F) f_b(x_2,\alpha_s,\mu_F)
\nonumber\\
&&\hskip2.0cm
\times \hat\sigma^{ab\to X}(sx_1x_2;\alpha_s,\mu_F,\mu_R).
\label{factorization}
\eea
Here $\mu_F$ and $\mu_R$ are the factorization and renormalization scales,
which are in principle arbitrary.  In practice, truncating
the cross section at a given order in perturbation theory
induces dependence on $\mu_F$ and $\mu_R$.

Although parton distributions are nonperturbative quantities which
must be measured experimentally at some short-distance scale $\mu$, 
their evolution with $\mu$ is governed by the DGLAP equation~\cite{DGLAP},
\be
{\del f_a(x,\mu) \over \del \ln\mu^2}
= {\alpha_s(\mu) \over2\pi} \int_x^1 {d\xi\over\xi}
 P_{ab}(x/\xi,\alpha_s(\mu)) f_b(\xi,\mu),
\label{DGLAPeq}
\ee
whose kernel is known through NNLO~\cite{MVVNNLO},
\be
 P_{ab}(x,\alpha_s) = 
 P_{ab}^{(0)}(x) 
+ {\alpha_s\over2\pi} P_{ab}^{(1)}(x) 
+ \left( {\alpha_s\over2\pi} \right)^2 \!\! P_{ab}^{(2)}(x) 
+ \Ord(\alpha_s^3).
\label{DGLAPkernel}
\ee
The partonic cross section can be expanded similarly
in powers of $\alpha_s$,
\bea 
&&\hat\sigma^{ab\to X}(\alpha_s,\mu_F,\mu_R)
\nonumber\\
&&=\ [\alpha_s(\mu_R)]^{n_\alpha}
\Biggl[ \hat\sigma^{(0)}
+ {\alpha_s(\mu_R)\over2\pi}\, \hat\sigma^{(1)}(\mu_F,\mu_R) 
\nonumber\\
&&+ \left( {\alpha_s(\mu_R)\over2\pi} \right)^2
\hat\sigma^{(2)}(\mu_F,\mu_R)
+ \Ord(\alpha_s^3) \Biggr],
\label{partoniccross}
\eea
where $n_\alpha$ depends on the process.
For typical collider processes, $\mu_R$ might be of order
100 GeV, for which $\alpha_s(\mu_R) \approx 0.1$. 
One might expect that the leading-order (LO), or Born level, 
terms in the expansion ($\hat\sigma^{(0)}$) would suffice to get 
a 10\% uncertainty. However, for hadron collider cross sections, 
corrections from the next-to-leading order (NLO) terms in 
the $\alpha_s$ expansion ($\hat\sigma^{(1)}$)
can increase the cross section by 30\% to 80\%.   
There are several reasons for the large corrections, 
some of which we shall discuss below.
Thus, LO predictions are only qualitative; quantitative predictions
require NLO corrections.  If a few percent precision is desired, 
then the next-to-next-to-leading order (NNLO) terms, may also be required.
Also one must be careful to describe the experimental setup (cuts, 
{\it etc.}) sufficiently accurately.

\subsection{Basic ingredients at fixed order}

\begin{figure}
\centerline{\epsfxsize 3.35 truein \epsfbox{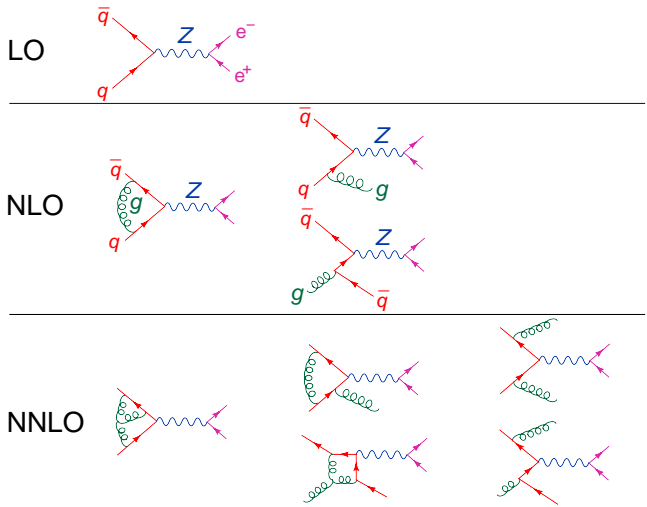}}
  \caption{Sample Feynman diagrams contributing to $Z$ boson production
at a hadron collider, at LO, NLO, and NNLO.  Only one diagram is shown
for each contributing amplitude, and some amplitudes are omitted.}
 \label{ZLONLONNLOfigure}
\end{figure}

What ingredients enter a perturbative QCD calculation at LO, NLO,
or NNLO?  First of all, various partonic scattering amplitudes are
required.  These amplitudes are illustrated in \fig{ZLONLONNLOfigure}
for one of the simplest processes, the inclusive production of a $Z$ boson
at a hadron collider, followed by $Z$ decay to an electron-positron pair. 
At LO, only tree amplitudes are needed.  In this example,
a single Feynman diagram contributes to $q\bar{q} \to Z \to e^+e^-$.
This diagram just needs to be squared, and convoluted with
the pdfs, while incorporating any experimental cuts on the final state
leptons.

At NLO, one-loop amplitudes contribute to virtual corrections;
for example, the one-loop correction to $q\bar{q} \to Z$.
The virtual corrections must be combined with real radiation;
{\it i.e.}, tree amplitudes having one additional parton in the final
state.  In the $Z$ example, the subprocesses are $q\bar{q} \to Zg$,
$qg\to Zq$ and $\bar{q}g\to Z\bar{q}$.
The virtual and real corrections are separately divergent in the
infrared (IR), which includes both soft and collinear regions.
Usually the IR divergences are regulated dimensionally, by letting the
number of spacetime dimensions be $D=4-2\e$ (with $\e<0$),
and expanding both virtual and real contributions in a singular 
Laurent expansion around $\e=0$.  There are $1/\e^2$ singularities
that cancel between virtual corrections and real corrections.
Some of the $1/\e$ singularities
also cancel this way; others, representing initial-state collinear 
singularities, are absorbed into a renormalization of the pdfs.
Ultraviolet poles are removed by coupling renormalization.
The finite remainder is then convoluted with the pdfs, as at LO.

At NNLO, there are three types of terms:
two-loop virtual corrections to the lowest-order process;
mixed virtual/real corrections from one-loop amplitudes with one additional
parton; and tree amplitudes with two additional partons,
as shown in \fig{ZLONLONNLOfigure}.
The IR cancellations are increasingly intricate, beginning now
at order $1/\e^4$.  

%%%%%%%%%%%%%%%%%%%%
%FIGURE

As the number of final-state partons in a process grows, 
the complexity of the theoretical computation increases,
at every order in $\alpha_s$, but the issue is particularly problematic
at NLO and NNLO.  Consider, as an example, the processes $pp \to\ n$ jets.
At LO, fast numerical programs allow the computation of
the $n$-jet final-state up to about 8 jets~\cite{BigTrees}, 
depending on the computing time available.  
At NLO, there are no complete results for more than
three jets.  In this case, the main limitation is the lack of knowledge of
the one-loop amplitudes for more than five external partons 
(except for one case, $gg \to gggg$).  
At NNLO, even the basic two-jet final state cannot yet be computed at NNLO.
Here the required amplitudes are all known, and the main obstacle has been
the integration over the singular final-state phase space. 
More interesting final states may include electroweak particles such
as $W$, $Z$, or Higgs bosons in addition to jets.  In each case,
the state-of the-art value of $n$ is the same or smaller as in the
$n$-jet case, if one counts each electroweak particle as replacing one jet.

\subsection{Singular phase-space integration}

Most of the recent advances in computing NNLO corrections to collider
processes have come from the development of methods for integrating
over singular regions of phase space where one or two partons are
``unresolved'', {\it i.e.}, are either soft or are collinear with
another, hard parton.  There have also been recent advances at NLO, 
in the context of matching parton showers with fixed-order computations,
which exploit an understanding of the singular phase-space structure.

At NLO, only one parton can be unresolved.  For example, consider
the final-state gluon in the tree-level amplitude for $q\bar{q} \to Z g$
in \fig{ZLONLONNLOfigure}.
There is a soft singularity when the gluon momentum vanishes, $k_g \to 0$.
This singularity can be interpreted classically, as radiation from the 
``accelerating'' quark color charges when the quark and antiquark
annihilate.  Thus soft radiation is associated with a pair of 
external partons.
There are also two collinear singularities, one in which the gluon
is collinear with the quark, from the Feynman diagram shown in 
\fig{ZLONLONNLOfigure}; and one in which the gluon
is collinear with the antiquark, from another Feynman diagram (not shown).

Various techniques have been developed for handling these 
singular integrals at NLO, and some of them are also being applied 
at NNLO.  There are four general categories of methods:
\begin{itemize}
\item{{\bf analytic}} --- Usually carried out on a case-by-case
  basis, for the simplest processes only, with (at best)
  very simple cuts.
\item{{\bf slicing}~\cite{Slicing}} --- Thin strips are excised
from the singular regions of phase space.  An approximate version
of the cross section can be integrated analytically in the strips.
Outside the strips, the integral is finite, so it can be performed
numerically, with generic experimental cuts. 
\item{{\bf subtraction}~\cite{ERT,Subtraction,FKS,CS,AntKosower,DGM}} --- 
One subtracts, over the entire phase space, a function mimicking the 
exact cross section in singular regions, 
but which can be integrated more easily, even in the presence of cuts.
The subtracted difference is integrated numerically.
\item{{\bf direct numerical integration}~\cite{GDGH4,DirectNum,FEHiP}} 
--- This approach can be carried out for a variety of processes, even at NNLO,
but most effectively after using sector decomposition~\cite{SectorDecomp}
to help separate the singularities.
\end{itemize}

\begin{figure}
\centerline{\epsfxsize 3.0 truein \epsfbox{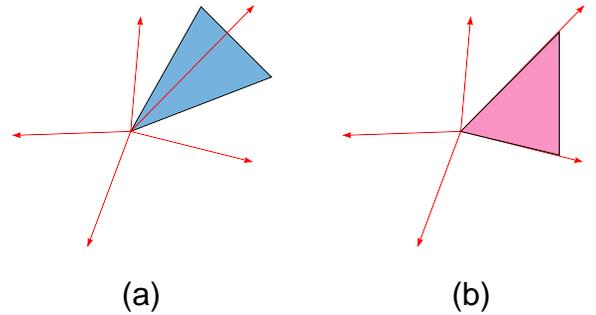}}
  \caption{Schematic depiction of two different types of subtraction
    methods.  Arrows represent momenta of hard external partons.
(a) The dipole subtraction method is built around collinear
singularities associated with individual partons.  
(b) The antenna subtraction method is built around soft
singularities associated with pairs of partons.}
\label{DipAntfigure}
\end{figure}

Within the category of subtraction methods, one can distinguish
two different ways of building up subtraction terms for general 
processes out of simpler building blocks.  The problem is that 
soft singularities connect all possible pairs of legs, while
collinear singularities are associated with individual legs.
Also, soft and collinear regions overlap, so any smooth subtraction
term should have a soft part and a collinear part.
\Fig{DipAntfigure}(a) sketches the building block for the
{\it dipole subtraction} method~\cite{CS}.
This function captures the collinear behavior near a
particular hard parton, plus part of the soft behavior connecting
that parton to other color-correlated partons.\footnote{%
The term ``dipole'' used here differs from the ``dipole shower'' used
in the Monte Carlo community, which is closer to an antenna pattern.}
\Fig{DipAntfigure}(b) illustrates the building block for the
{\it antenna subtraction} method~\cite{AntKosower}.
It captures the soft behavior associated with a pair of color-connected
partons, plus part of the collinear behavior near each of the members
of the pair.

%%%%%%%%%%%%%%%%%%%%
%FIGURE

\subsection{Parton showers and NLO matching}

Parton showers represent an approximation to the soft and collinear
radiation pattern, a resummation of leading logarithms (at present),
which can be implemented probabilistically.   Parton showers
are a key part of Monte Carlo simulation programs
such as {\tt PYTHIA}~\cite{PYTHIA} and {\tt HERWIG}~\cite{HERWIG},
which produce hadron-level events and are essential to experimental
analyses.  To improve the accuracy of parton-shower Monte Carlo programs,
it is important to match them to fixed-order results.  LO matching is
an important subject which I cannot do justice to here; see however,
ref.~\cite{LOMatching} for a recent comparative study of different
approaches.  

For even better accuracy, though, NLO matching is necessary.
The program {\tt MC@NLO}~\cite{MCNLO} was the first to accomplish
NLO matching for a variety of different processes.  It operates within the 
{\tt HERWIG} Monte Carlo, which uses an angular-ordered shower.  
Implementing additional processes in {\tt MC@NLO} has been nontrivial.
An important issue is to avoid the double-counting of emissions,
between ({\it i}) the first step of the shower, and 
({\it ii}) the exact radiation pattern in QCD, from which subtraction 
terms have been removed in the course of the NLO computation.
If the shower radiation pattern differs from the form of the subtraction
terms, a correction is necessary.

Recently there have been some advances in achieving NLO matching in a
more process-independent way.  The {\tt POWHEG} method~\cite{POWHEG},
in contrast to {\tt HERWIG}, generates the hardest parton emission first,
and works well with $p_{\rm T}$-ordered showers.  Recently
it has been formulated for a general NLO subtraction method,
and in particular for the Frixione-Kunszt-Signer~\cite{FKS} and
dipole~\cite{CS} methods, for generic processes~\cite{POWHEGFNO}.

It has also been recognized~\cite{NSshower} that subtraction methods
automatically supply radiation patterns that are correct in the soft
and collinear limits, and hence can be used to construct a parton shower.
In this case the double-counting problem is solved automatically, because the
first step of the shower coincides with the NLO subtraction function.
Two independent implementations of a shower based on
dipole subtraction have appeared very recently~\cite{CSshower},
as well as one ({\tt VINCIA}) based on antenna subtraction~\cite{VINCIA}.
With the recent flourishing of different parton-shower algorithms and
NLO matching routines, it will be very interesting to compare their
outputs for benchmark processes over the next year or two.

\subsection{Subtraction at NNLO}

Let us return now to the status of fixed-order results.
The NLO technique that has been carried out for the widest variety
of collider processes is the dipole subtraction method.  
Programs such as {\tt MCFM}~\cite{MCFM} and {\tt NLOJET++}~\cite{NLOJET}
cover a variety of hadron collider, $ep$ collider and $e^+e^-$ processes,
limited mainly by the availability of virtual corrections (one-loop
amplitudes).  The method has been generalized to handle massive
final-state partons (such as top quarks)~\cite{CDST};
and a fully automated version of the method in the massless case has appeared 
very recently~\cite{GK}.  

It is natural to try to generalize this method to NNLO.  Such a
generalization is highly non-trivial, because there are now
several different types of singular phase-space regions, for the 
contributions with two additional radiated partons, exemplified
by the rightmost terms on the NNLO line in \fig{ZLONLONNLOfigure}.
There can be singularities when both additional partons are soft,
when one is soft and the other collinear; when both are collinear
with a third parton; and when there are two independent collinear pairs.
These different regions overlap with each other.
Nevertheless, progress has been made in constructing dipole-type 
subtraction terms for $e^+e^-$ annihilation to jets~\cite{NNLOCS}.
The method has also been adapted to give NNLO results
for the inclusive production of a Higgs boson via gluon fusion 
at hadron colliders~\cite{CG}.   

The most complex process that has been treated to date at NNLO
is that of three-jet event shapes in $e^+e^-$ annihilation, 
such as the thrust distribution~\cite{GDGGHThrust} to
be described further in the next section.  These results
relied on building antenna-type subtractions to handle all the singular
regions, and evaluating their phase-space integrals~\cite{GDGH4,GDGG}.

\subsection{Sector decomposition at NNLO}

\begin{figure}
\centerline{\epsfxsize 3.4 truein \epsfbox{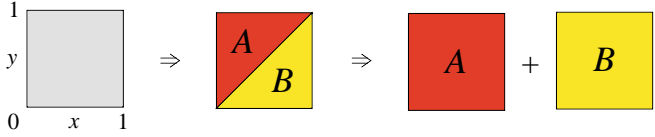}}
  \caption{Sector decomposition splits and remaps integration regions 
in order to expose multi-variable singularities.}
\label{sectorfigure}
\end{figure}

Iterated sector decomposition is a strategy for doing singular
integrals by partitioning the integration region, and then remapping
it, in order to make the singularities one-dimensional.
A very simple example of the procedure starts with the integral
\be
I = \int_0^1 \int_0^1 dx dy { x^\e y^\e \over (x+y)^2 } \,.
\label{sectorex}
\ee
Integrals like $I$ are encountered in the NNLO corrections 
to vector boson production (for example), 
from the interference of two different initial-state radiation 
graphs in which a parton, radiated from either incoming line,
splits into two additional partons.  Although $I$ is not singular
as either $x$ or $y$ approaches zero with the other variable held fixed, 
there is a singularity as $x=y \to 0$.  This singularity can be exposed
by splitting the unit square into sectors $A$ and $B$ shown in 
\fig{sectorfigure}, and then remapping region $A$ back to the unit square
using $x'=x$, $1-y'=(1-y)/(1-x)$, and $B$ back to the unit square
using $y'=y$, $1-x'=(1-x)/(1-y)$.  These transformations map the
$x=y \to 0$ singularity onto one variable, $x'\to0$ or $y'\to0$,
depending on the sector.  This technique was first
applied to multi-loop integrals~\cite{SectorDecomp},
and later to phase-space integrals~\cite{GDGH4,DirectNum,FEHiP}.
Several iterations and many sectors may be required for state-of-the-art 
NNLO results.  The expansion in $\e$ for one-dimensional singularities is
straightforward, involving standard ``plus'' distributions.
Hence arbitrary observables can be integrated over phase space.

%%%%%%%%%%%%%%%%%%%%
%FIGURE

\subsection{Soft-gluon resummation}

In particular kinematic regions, fixed-order perturbation
theory breaks down.  This breakdown can be due to a mismatch
between the kinematics of virtual and real corrections,
enhanced by the strength of soft-gluon emission.
For example, in $Z$ production by $q\bar{q}$ annihilation, 
the transverse momentum of the $Z$, $q_{\rm T}(Z)$, vanishes for 
all virtual corrections, but is nonzero for real corrections, 
which are enhanced for small $q_{\rm T}(Z)$.  The leading
behavior at $L$ loops is
\be
{d\hat\sigma\over dq_{\rm T}^2(Z)} \sim 
(C_F\alpha_s)^L { \ln^{2L-1} q_{\rm T}^2(Z) \over q_{\rm T}^2(Z) }
 + \cdots \,
\label{qTsing}
\ee
plus terms with fewer logarithms.  Here $C_F = 4/3$
is the quark color charge.  These large corrections
imply that {\it transverse-momentum resummation}~\cite{pTResum} 
is required for $q_{\rm T}(Z) \ll M_Z$.  

Distributions in the threshold variable $z \equiv M_Z^2/s_{q\bar{q}}$
behave similarly.  Virtual corrections have $z=1$, while real corrections
have $z<1$, and are enhanced as $z\to1$, with leading behavior,
\be
{d\hat\sigma\over dz} \sim 
(C_F\alpha_s)^L \biggl[ { \ln^{2L-1}(1-z) \over 1-z } \biggr]_+ + \cdots \,.
\label{zsing}
\ee
Such singular distributions can give rise to large corrections
to inclusive production cross sections if they are convoluted
with steeply falling parton distributions, necessitating 
{\it threshold resummation}~\cite{ThrResum}.

The kinematics are similar for production of the Higgs boson via
gluon fusion, except that the color charge for gluons, $C_A = 3$,
is much larger than for quarks, and the gluon pdfs are falling much 
faster than the quark pdfs in the relevant $x$ range, $x\sim 10^{-2}$.
Hence both $p_{\rm T}$ and threshold resummation effects are much 
larger in this case than for vector boson production.

%%%%%%%%%%%%%%%%%%%%%%%%%%%%%%%%%%%%%%

\section{An $e^+e^-$ application -- thrust at NNLO}
\label{ThrustSection}

Next I turn to some recent applications of these theoretical
techniques to collider processes, beginning with one from
$e^+e^-$ annihilation.  The thrust~\cite{ThrustDef} is a 
classic infrared-safe $e^+e^-$ event-shape observable,
defined by 
\be
T = {{\rm max} \atop \hat{n}}\ 
 { \sum_i |\vec{p}_i\cdot \hat{n} | \over \sum_i |\vec{p}_i | } \,.
\label{thrustdef}
\ee
The sum is over the final-state hadrons (or partons,
in a perturbative calculation) with spatial momenta $\vec{p}_i$,
and $\hat{n}$ is a unit vector, varied over all directions.
There is a wealth of data to compare to, for $\sqrt{s}$ 
from 14 to 206~GeV, 
with the best statistics gathered on the $Z$ pole.
In the extraction of $\alpha_s(M_Z)$ from a fit of NLO QCD
predictions to $e^+e^-$ event shapes, it is a long-standing problem
(see {\it e.g.} ref.~\cite{ALEPH04}) that the central value depends 
on the observable used, and on the range in that observable used 
in the fit.  The error is dominated by the truncation of the perturbative
series at NLO.

Finally, 27 years after the first NLO event-shape results~\cite{ERT}, 
the first NNLO results have appeared this year~\cite{GDGGHThrust}.
At the $Z$ pole, the effect of the NNLO corrections is to 
increase the NLO thrust distribution
by about 15--20\% in the range $0.04 < 1-T < 0.33$.
(The two-jet region, $1-T < 0.03$, requires resummation.
The region $1-T > 0.33$ cannot be produced by $q\bar{q}g$ final states,
so it is also less perturbatively stable.)
The relative uncertainty in the perturbative prediction is reduced
by about 30--40\% with respect to NLO.

The increase in the thrust distribution should lead to a somewhat
smaller $\alpha_s(M_Z)$; but at NLO the thrust led to a larger than
average value, $\alpha_s(M_Z)\approx 0.126$.
A new, more precise value of $\alpha_s(M_Z)$ requires experimental
re-analysis, incorporating:
\begin{enumerate}
\item a resummation of large logarithms for $1-T \to 0$,
\item an analysis of power corrections of the form $\Lambda_{\rm QCD}/Q$
using data off the $Z$ pole,
\item other event-shape observables, which have also been computed 
(very recently) at NNLO~\cite{GDGGHOther}.
\end{enumerate}
Indeed, the first NNLO determination of $\alpha_s$ from event-shape
variables, using ALEPH data for six different variables
(including thrust) from several center-of-mass energies,
has just been reported (post LP07)~\cite{DGDGGHOS},
\bea
\alpha_s(M_Z) &=& 0.1240 \pm 0.0008_{\rm stat}
                       \pm 0.0010_{\rm exp}
\nonumber\\
&&\hskip1.0cm
                       \pm \, 0.0011_{\rm had}
                       \pm 0.0029_{\rm theo}\,.~~~~~
\label{NNLOevsalpha}
\eea
The result is a bit higher than the world average~(\ref{alphasworld04}).
The perturbative uncertainty, ``theo'', has been cut roughly in half
with respect to NLO.  Further analyses, incorporating resummation 
and power corrections, are eagerly awaited! 

%%%%%%%%%%%%%%%%%%%%%%%%%%%%%%%%%

\section{Higgs production at hadron colliders}
\label{HiggsSection}

\subsection{Gluon fusion total cross section}

Let's begin the discussion of hadron collider applications with
Higgs boson production and decay.  These results are
of great phenomenological importance for the Higgs search at the
Tevatron, and particularly at the LHC.  The gluon-fusion channel,
$gg \to H$, via a top quark loop, represents one of the simplest
final states, apart from vector boson production, so the theory 
has already been pushed to rather high order.  
At the same time, the process illustrates methods of analysis 
that should eventually be applied to more complex states.

In the most likely mass range for the Standard Model Higgs boson, from 
114~GeV to about 200 GeV, gluon fusion dominates the total
production cross section $\sigma_H$.  
It has been known since the early 1990s that the NLO corrections 
to $\sigma_H$ were huge, increasing it by roughly $80\%$~\cite{NLOHiggs}.
Because the lowest-order process proceeds by a loop diagram,
the NLO corrections already require two-loop integrals.
To go to NNLO, the large-$m_t$ approximation has been used.
This approximation shrinks the top-quark triangle to a point,
replacing it by an effective operator, 
$H \, G^a_{\mu\nu} G^{\mu\nu\,a}$~\cite{HggOperator},
with coefficient $C_H$.  It
reduces the number of loops and the number of mass scales in the problem
by one.  This makes it feasible to perform all the NNLO computations, 
including the phase-space integrals, analytically for the case of 
inclusive production~\cite{NNLOHiggs}.  Threshold logarithms,
of the form 
\be
{d\hat{\sigma}_H\over dz} \sim 
(C_A\alpha_s)^L \biggl[ { \ln^{k}(1-z) \over 1-z } \biggr]_+ \,,
\label{zsingH}
\ee
with $k=0,1,\ldots 2L-1$, play a big role in the large positive 
corrections.  They have been resummed to next-to-next-to-leading 
logarithmic (NNLL) accuracy~\cite{CDFG}.
Nevertheless, a sizable uncertainty of order 10--15\% remained 
in the prediction.  

\begin{figure}
\centerline{\epsfxsize 3.4 truein \epsfbox{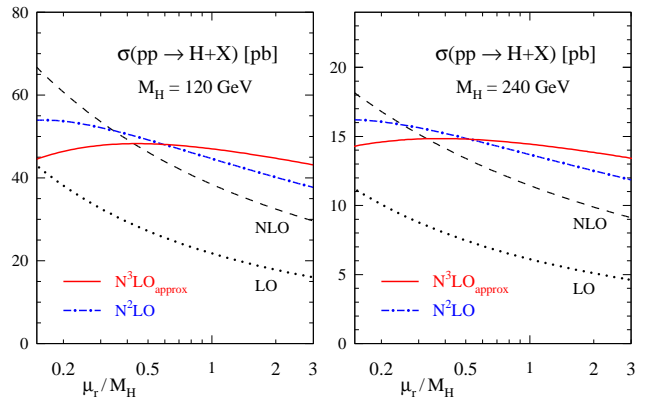}}
  \caption{Variation of the total Higgs production cross section
with respect to the renormalization scale $\mu_r$, at various
levels of approximation.  Figure from ref.~\cite{MV05}.}
\label{hmuvarfigure}
\end{figure}

Some progress has been made on reducing this uncertainty.
As a spinoff from the computation of the NNLO DGLAP kernel
for gluon evolution, $P_{gg}^{(2)}(x)$~\cite{MVVNNLO},
Moch and Vogt~\cite{MV05} were able to extract the leading singular
terms~(\ref{zsingH}) at three loops, $L=3$ and $k=0,1,2,3,4,5$.
They defined an approximation N$^3$LO$_{\rm approx}$ which
is missing just the $\delta(1-z)$ term, plus all terms that
are nonsingular as $z\to1$.  They also employ the N$^3$LO corrections
to the coefficient $C_H$~\cite{CKS}.  The results are shown
in \fig{hmuvarfigure}.  The renormalization-scale dependence
is now stabilized near the Higgs mass, with a residual uncertainty
of order 5\%.  More recently, a similar analysis has been carried
out at order ``N$^4$LO$_{\rm approx}$'' with the six leading terms
in \eqn{zsingH} at that order~\cite{RSvN}.  The results change by
about 2\% with respect to the N$^3$LO$_{\rm approx}$ results
shown in the figure.  The total Higgs production cross section
is now becoming a precision observable.

%%%%%%%%%%%%%%%%%%%%
%FIGURE

\subsection{Gluon fusion with decay to two photons}

Unfortunately, the total Higgs production cross section cannot
be observed experimentally.  For any final state, various
experimental cuts have to be imposed.
For example, for the two-photon final state from the
decay $H\to\gamma\gamma$ there are cuts
on the photons' transverse momentum, rapidity, and isolation
(with respect to hadronic energy).  
These cuts can now be mimicked at the parton level at NNLO,
for gluon fusion production (in the large-$m_t$ limit)
followed by $H\to\gamma\gamma$. 
(The same is now also true for the decay mode
$H\to WW\to \ell\nu\ell\nu$~\cite{HWWNNLO,CG}.)
Two independent programs are available for the
$\gamma\gamma$ final state.
The program {\tt FEHiP}~\cite{FEHiP}
uses direct numerical integration, following sector decomposition 
for the real corrections.   The more recent program
{\tt HNNLO}~\cite{CG} employs a subtraction method, based
on the fact that at nonzero Higgs $q_{\rm T}$ the problem is 
really a NLO calculation;
plus a knowledge, based on resummation studies, 
of the universal behavior of the $q_{\rm T}$ distribution as
$q_{\rm T} \to 0$.

\begin{figure}
\centerline{\epsfxsize 3.4 truein \epsfbox{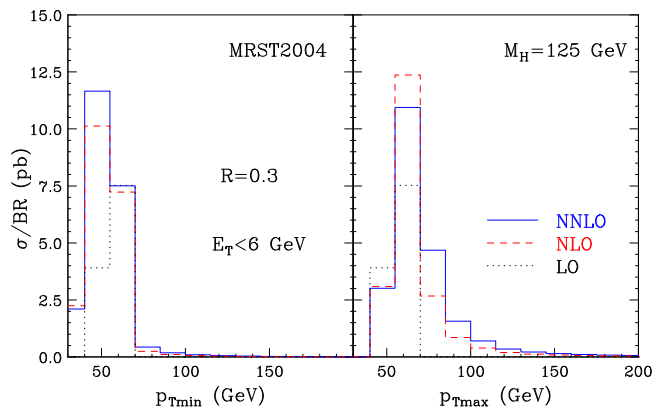}}
  \caption{Distributions of transverse momenta $p_{\rm T}$
for the minimum and maximum $p_{\rm T}$ photons for
Higgs production at the LHC, followed by the decay $H\to \gamma\gamma$, 
at LO, NLO and NNLO. Figure from ref.~\cite{CG}.}
\label{HNNLOpTfigure}
\end{figure}

For regions of final-state phase space that are accessible at LO,
both programs show good perturbative stability in going from NLO to NNLO.
However, the behavior near boundaries of the LO-accessible region
is more unstable, again reflecting the breakdown of fixed-order
perturbation theory.  One such boundary occurs at a large rapidity
difference between the two photons~\cite{FEHiP}.  At $q_{\rm T} = 0$,
the photon $p_{\rm T}$ cuts and fixed Higgs mass $m_H$ put a bound on how 
forward and backward the decay photons can go.
More obvious boundaries occur in the transverse momenta of 
the two photons, $p_{\rm Tmin}$ and $p_{\rm Tmax}$, which
each have to be less than $m_H/2$ at LO.  One can see from
\fig{HNNLOpTfigure} that the $p_{\rm Tmax}$ distribution gets
significantly stiffer, and the $p_{\rm Tmin}$ distribution gets
softer, from NLO to NNLO~\cite{CG}.  
(The distributions are forced to be identical at LO.)

%%%%%%%%%%%%%%%%%%%%
%FIGURE

\subsection{The two-photon background}

In searching for a bump from the Higgs boson in the invariant-mass 
of photon pairs, it is also useful to know the characteristics
of the continuum QCD background.  The $\gamma\gamma$
background has many components, including:
\begin{itemize}
\item electrons, positrons and hadrons faking photons
\item photons arising from copious $\pi^0$ decays and 
electron or positron bremsstrahlung
\item photons arising from fragmentation --- 
radiation at small transverse momentum with respect to a jet
\item hard QCD radiation of photons.
\end{itemize}
The first three categories of backgrounds can be suppressed
fairly effectively by photon isolation cuts.

Direct production of photon pairs in hard QCD begins at LO with
the quark-annihilation process $q\bar{q}\to\gamma\gamma$.
The process $qg \to \gamma\gamma q$ enters at NLO.  It
is heavily enhanced, by the large gluon pdf at small $x$,
and by a final-state collinear singularity between each photon 
and the outgoing quark.  (At very small angles, the singularity 
is absorbed into the fragmentation contribution; it is also suppressed
by isolation cuts.)  At NNLO, the process $gg\to\gamma\gamma$,
mediated by a virtual quark loop, enters for the first time.
It is significant, however, because the gluon pdf enters twice.
NLO programs for the inclusive di-photon background, 
including fragmentation effects,
were constructed in the late 1990's~\cite{NLOdiphoton,DIPHOX}.
The contribution from the $gg\to\gamma\gamma$ subprocess
at {\it its} next-to-leading order in the pdf-enhanced
channel $gg\to \gamma\gamma(g)$ was added a few years
later~\cite{BDS02}.  This year the quark-box contribution to
$qg\to\gamma\gamma q$ was included, as well as a resummation
of the transverse momentum $q_{\rm T}$ of the di-photon pair at
NNLL accuracy~\cite{BBNY07a,BBNY07b}. 

\begin{figure}
\centerline{\epsfxsize 1.8 truein \epsfbox{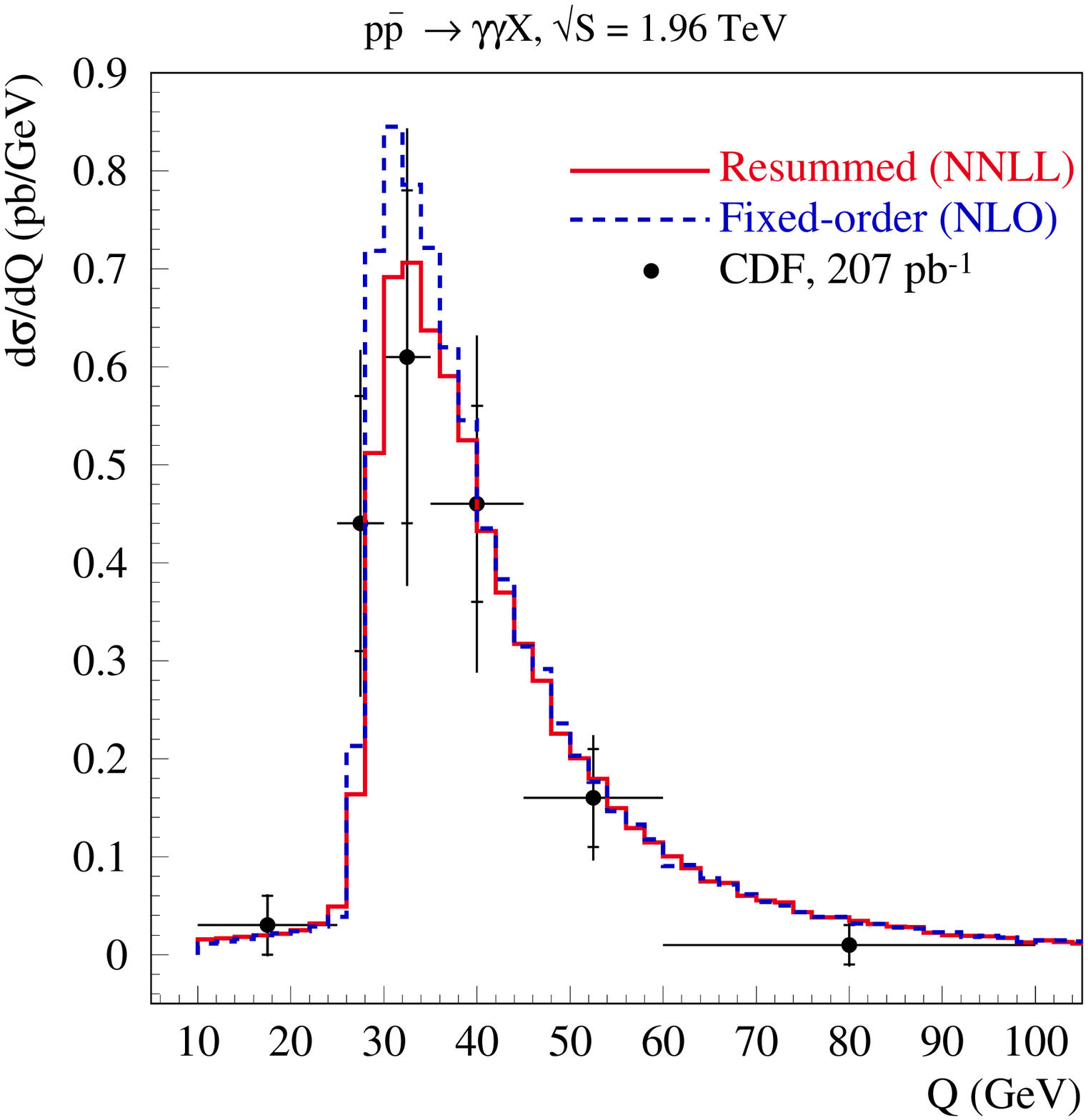}\ 
\epsfxsize 1.8 truein\epsfbox{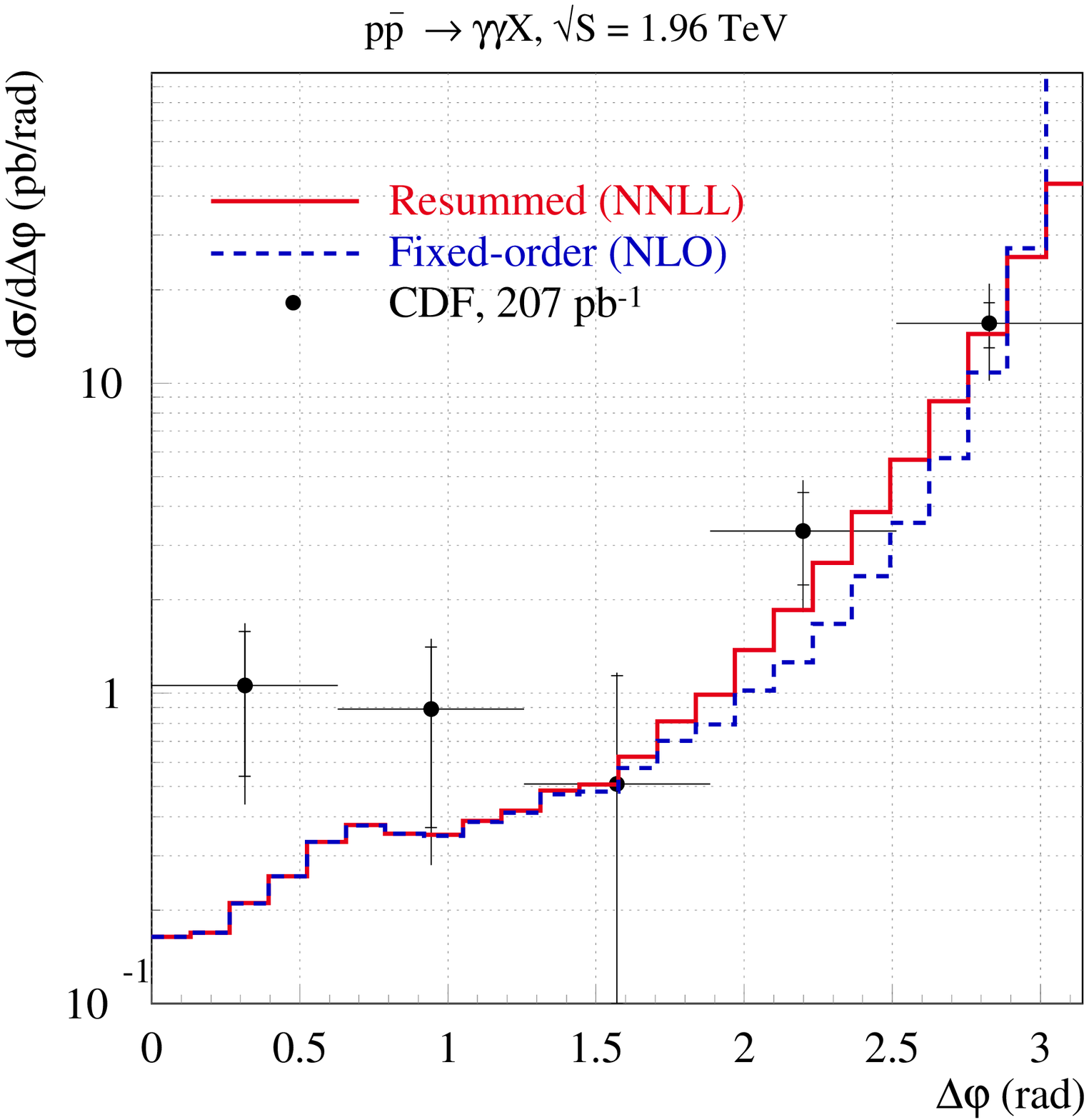}}
\caption{Distributions in invariant mass $Q$ and azimuthal
angle separation $\Delta\phi$, for pairs of photons produced
at the Tevatron. Figure from ref.~\cite{BBNY07b}.}
\label{ggfigure}
\end{figure}

CDF has presented 207 pb$^{-1}$ of data~\cite{CDFdiphoton}
on pairs of photons produced at the Tevatron
with $p_{\rm T}^\gamma > 14$ GeV and di-photon invariant masses 
ranging from 10 to 100 GeV.  This range is useful for assessing
the hard QCD background to the Higgs search at the LHC.
If one scales a di-photon invariant mass of, say, 150 GeV,
and typical photon $p_{\rm T}$ cut of 40 GeV,
down by the ratio of beam energies between the LHC and the Tevatron
(seven), one gets a di-photon invariant mass of order 20 GeV,
and $p_{\rm T}$ cut of 6 GeV, not too far from the CDF cuts.
So similar ranges of $x$ values for the pdfs are being sampled.
\Fig{ggfigure} shows the CDF data, in comparison with the
NLO plus NNLL-resummed predictions~\cite{BBNY07b}.
The agreement is generally quite good, within the available
statistics, except for the region of small $\Delta\phi$.
Significant contributions to the small $\Delta\phi$ region
can arise when a single parton yields both photons,
{\it e.g.} $gq\to gq\gamma$, with hard photon radiation off the final-state
quark, followed by the fragmentation of that quark to a second photon.  
These fragmentation contributions are NLO
(order $\alpha^2 \alpha_s$), and are not included in the computation
of ref.~\cite{BBNY07b}.  They have been included in the program
{\tt DIPHOX}~\cite{DIPHOX}, which does fit the CDF data at small
$\Delta\phi$.  The small $\Delta\phi$ region will probably 
not be too important for the Higgs search; such large boosts
are kinematically disfavored.

%%%%%%%%%%%%%%%%%%%%
%FIGURE

In general, then, the di-photon background
at the LHC seems to be in relatively good shape; although
a computation of the $qg\to\gamma\gamma g$ channel at {\it its} NLO,
leading eventually to a computation of $q\bar{q}\to\gamma\gamma(gg)$
at NNLO, would certainly be welcome.

\subsection{Vector boson fusion}

\begin{figure}
\centerline{\epsfxsize 3.4 truein \epsfbox{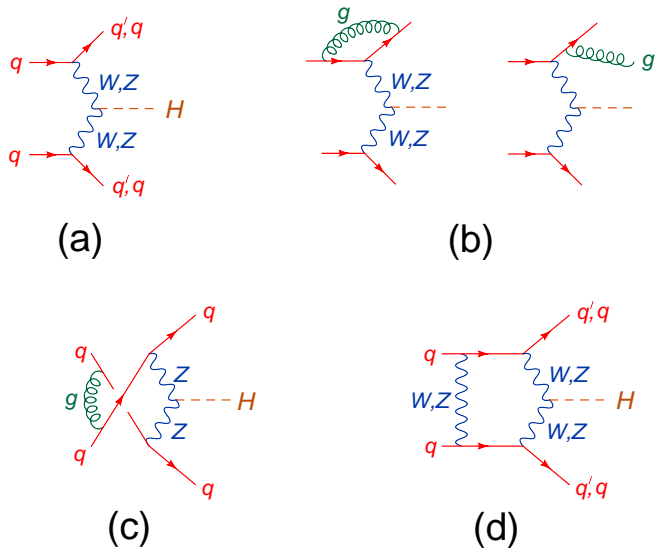}}
\caption{(a) LO diagram for weak boson fusion. (b) Most important
NLO QCD corrections. (c) Example of an additional NLO QCD correction.
(d) Sample NLO electroweak correction.}
\label{WBFfigure}
\end{figure}

The second largest Higgs boson production mechanism, weak boson
fusion (WBF), $qq \to qqH$, features a pair of forward tagging jets,
as shown in \fig{WBFfigure}(a).
The NLO QCD corrections that dominate in the forward limit
are sketched in \fig{WBFfigure}(b).  They only involve one quark
line at a time, and have been known for a while to be quite modest, 
only 5\% or so~\cite{NLOWBF,FZWBF}.  Recently, the complete set
of QCD corrections to WBF were computed,
as well as the one-loop electroweak corrections to all
channels~\cite{CDD07}.   An example of an additional QCD
correction is shown in \fig{WBFfigure}(c); it can interfere
with the graph in \fig{WBFfigure}(a) if the two quarks are identical.
There are also $s$-channel annihilation graphs.  However,
these contributions are kinematically disfavored after typical WBF cuts
(emphasizing forward jets) are imposed.  Ref.~\cite{CDD07} confirms 
that these contributions are tiny.   On the other hand, 
the one-loop electroweak corrections
are found to be sizable and negative, of order $-7$\%, 
so they must be included along with the QCD corrections. 

%%%%%%%%%%%%%%%%%%%%
%FIGURE

\begin{figure}
\centerline{\epsfxsize 3.4 truein \epsfbox{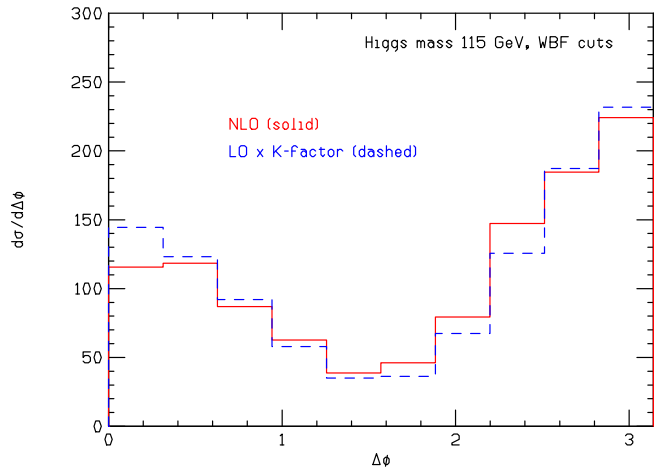}}
\caption{ Distribution in azimuthal separation $\Delta\phi$ for
tagging jets in the gluon-fusion-plus-two-jets process,
after applying WBF cuts, at LO (multiplied by the NLO K-factor),
and at NLO.  Figure from ref.~\cite{CEZ06}.}
\label{wbfbkgddelphifigure}
\end{figure}

Higgs boson production via WBF has various backgrounds, of course,
depending in general on the decay mode.
It also has a source of ``background''
that is independent of the decay mode, coming from the production 
of a Higgs boson via the gluon fusion subprocess ($Hgg$ interaction),
plus the radiation of at least two more jets.  
This ``background'' could in principle impact studies of the WBF
production mechanism.  While most of the gluon-fusion-plus-two-jets
background is eliminated by WBF cuts, it is important to understand
how various distributions for this subprocess are affected by
higher-order QCD corrections.  Recently, gluon-fusion-plus-two-jets
was computed at NLO, in the large-$m_t$ limit~\cite{CEZ06}.
The computation employed a semi-numerical evaluation of the one-loop
virtual corrections~\cite{EGZ05}.  The overall rate for this subprocess,
with typical WBF cuts, increases by 30\% in going from LO to NLO.
This increase is much less than that for the inclusive gluon-fusion
process, but still significant.  The azimuthal separation $\Delta\phi$
of the two tagging jets is an incisive probe of the WBF production
mechanism~\cite{FZWBF}.
\Fig{wbfbkgddelphifigure} shows how this distribution
changes in going from LO to NLO, in the gluon-fusion-plus-two-jets
``background'' subprocess.  The normalized distribution is 
fairly stable, only flattening slightly.

%%%%%%%%%%%%%%%%%%%%
%FIGURE

%%%%%%%%%%%%%%%%%%%%%%%%%%%%%%%%%

\section{Jets}
\label{JetsSection}

\subsection{Jet definitions}

Jets are by far the most copious high-transverse-momentum
objects produced at hadron colliders. They are common as well to almost all
$ep$ final states and a majority of $e^+e^-$ final states at high energy.
Hence a thorough understanding of their production rates and properties 
in various regimes is highly desirable.

There are two popular classes of algorithms for defining jets.
At $e^+e^-$ colliders, {\it cluster} algorithms have traditionally been
used~\cite{Cluster}.  A ``distance'' metric is defined between pairs 
of particles,
which vanishes when the two particles are collinear or one is soft.
The pair with the smallest distance is clustered into a proto-jet,
and the process is iterated, until all proto-jets are separated
by more than a specified distance, called the jet resolution parameter $y$.
This ``bottom up'' procedure automatically assigns every particle to a jet.
It is {\it infrared-safe} at the parton level:  Adding an arbitrarily 
soft gluon, or splitting one parton into two very collinear partons,
does not change any of the subsequent clustering steps, so the
final set of jets remains the same.  This means that jet rates
can be calculated (in principle) to any order in perturbation theory.

Events at hadron colliders typically contain, in addition to a hard 
partonic scattering process, an ``underlying event'', consisting
of forward beam remnants and particles produced more centrally by  
interactions of spectator partons with each other and with partons 
from the hard process.
The $k_{\rm T}$ algorithm~\cite{kT} is a cluster-style algorithm
that has been adapted for use at both $ep$ and hadron colliders.  
It clusters particles that are ``closer'' to the beam axis into a beam jet,
rather than into one of the other jets.  However, some energy from the 
underlying event will still be associated with non-beam jets, and
quantifying this is an issue, although it can be done~\cite{CDFkT}.

Traditionally, hadron collider jet algorithms have been based 
instead on {\it cones}.
Cone jets are circles of fixed radius $R$ in the plane of pseudorapidity 
$\eta$ and azimuthal angle $\phi$ containing a set of final-state 
hadrons, or a set of calorimeter towers, with total transverse momentum 
$p_{\rm T}^{\rm jet}$ above a particular lower threshold.
There are also rules for splitting and merging nearby cones 
in order to arrive at a stable configuration.
Usually the cones are ``seeded''; {\it i.e.}, there is a prescription for
specifying which circle centers will be considered first.  The seeds
might be towers with transverse momentum above a seed threshold.
Minimum-bias events distribute energy uniformly in the $\eta$-$\phi$
plane, making for a simple prescription for correcting for the
underlying event in a cone algorithm.  See {\it e.g.} 
refs.~\cite{CHS,EHHLT} for more details.  

%%%%%%%%%%%%%%%%%%%%
%FIGURE
\begin{figure}
\centerline{\epsfxsize 3.2 truein \epsfbox{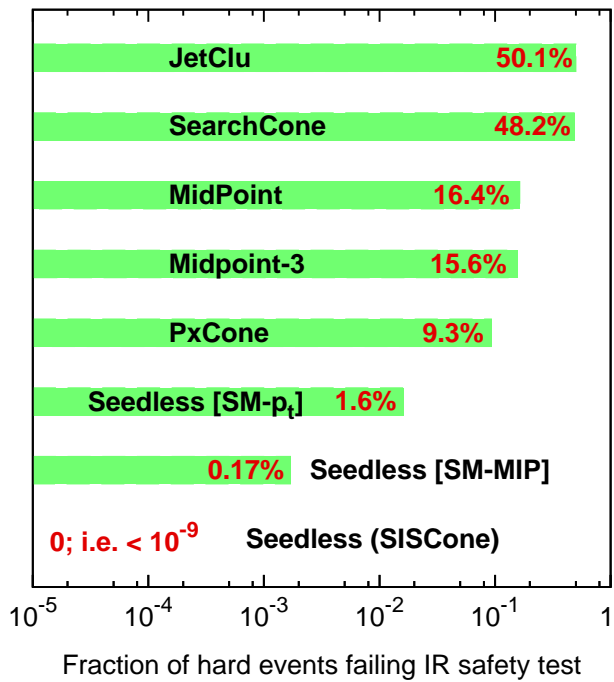}}
  \caption{Fraction of hard events that fail an infrared-safety test, 
for various cone algorithms.  Figure from ref.~\cite{SIScone}.}
\label{SISfigure}
\end{figure}

Unfortunately, for seeded cones there is a danger of infrared unsafety:  
The final jet configuration can sometimes change by a lot, 
with the addition of an arbitrarily soft gluon.
Introducing a midpoint seed between all the previous seeds can
reduce the problem, but not eliminate it.  In the past year, 
a new, computationally practical, seedless cone algorithm, SIScone, 
was invented~\cite{SIScone}.  This algorithm is infrared safe, as shown in
\fig{SISfigure}.   The figure also indicates the fraction of hard events 
that fail an infrared-safety test, for some other popular cone algorithms.

Packages are now available that give the user the flexibility to reconstruct
jets with a variety of algorithms and parameters, and to do comparisons
of jet properties between different algorithms~\cite{SIScone,EHHLT,SpartyJet}.

\subsection{Jet substructure}

To what extent can the identities of underlying partons 
be deduced from properties of the jets they produce?
In particular, can we distinguish light quark jets from gluon jets?
It is not possible to do this event by event; however, it is possible
to do it statistically, by studying how energy is distributed within 
the jet.  In more detail, event kinematics are used to select 
gluon-rich or gluon-depleted samples of jets.  For these samples,
one can measure the distributions of the jet-shape function $\Psi(r/R)$
--- the fraction of energy for a jet with cone size $R$ that is
found in a smaller cone with $r<R$.  If the distributions differ
for the gluon-rich and gluon-depleted samples, then one could use 
the jet-shape distributions to separate quark and gluon jets, 
statistically.  (Similar studies have been carried out in the past
using sub-jet multiplicities, in $e^+e^-$~\cite{SubJetee}, 
$ep$~\cite{SubJetep} and $p\bar{p}$~\cite{SubJetpp} collisions.)

In $p\bar{p}$ collisions, CDF studied the dependence of the jet-shape
distribution on the transverse momentum of jets reconstructed with the
midpoint algorithm ($R=0.7$)~\cite{CDFjetshape05}.  
Higher transverse-momentum jets are gluon-depleted, in comparison with 
lower transverse-momentum jets.  Good agreement was found with
PYTHIA~\cite{PYTHIA}, for transverse momenta ranging from
50~GeV (73\% gluon, 27\% quark, according to PYTHIA; 
$\langle\Psi(0.3/R)\rangle = 0.705\pm0.015$)
all the way up to 350~GeV 
(20\% gluon, 80\% quark; 
$\langle\Psi(0.3/R)\rangle = 0.93\pm0.02$).

\begin{figure}
\centerline{\epsfxsize 3.2 truein \epsfbox{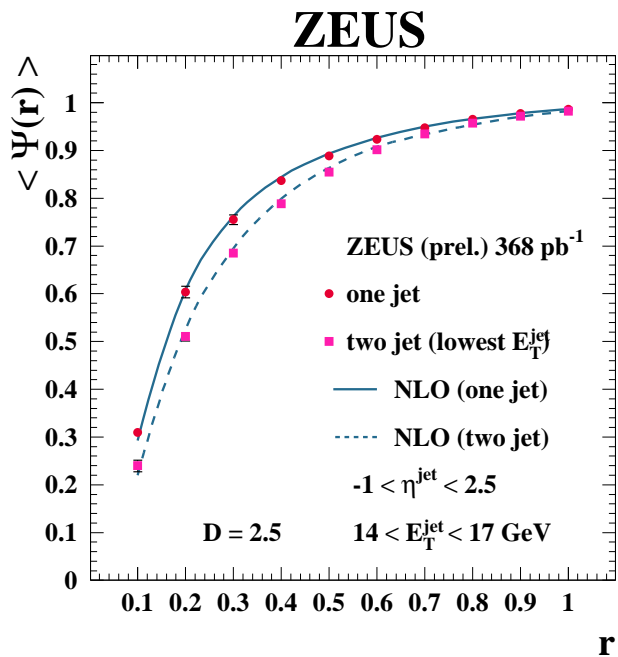}}
  \caption{Jet-shape distribution in $ep$ collisions for jets in
one-jet events, {\it vs.} the lowest $p_{\rm T}$ jet in two-jet events,
at the same jet $p_{\rm T}$.  Figure from ref.~\cite{ZEUSjetshape07}.}
\label{ZEUSJetShapefigure}
\end{figure}

Recently, ZEUS studied the jet-shape distribution in $ep$
collisions, with 368~pb$^{-1}$ of data, using two-jet events
and taking the lowest transverse-momentum jet in order to select a 
gluon-enriched sample~\cite{ZEUSjetshape07}.
\Fig{ZEUSJetShapefigure} shows the ZEUS data,
using the $k_{\rm T}$ algorithm with resolution parameter $D=2.5$
and jet transverse momenta in the narrow range between 14 and 17~GeV.
The agreement with NLO predictions (from the program {\tt DISENT}\cite{CS}) 
is excellent, to within a percent or so.
Thus the substructure of jets in generic events in both $p\bar{p}$
and $ep$ collisions is understood very well now.

%%%%%%%%%%%%%%%%%%%%
%FIGURE

\subsection{Inclusive jet rates}

The production rate for very high transverse-momentum jets
at the Tevatron provides a direct test of QCD interactions
at the shortest possible distance scales.  Of course, the rate
also depends on the parton distribution functions, in particular
the large-$x$ gluon distribution.
Because jet production rates are rapidly-falling functions 
of $p_{\rm T}$, they are very sensitive to the jet energy scale 
calibration, as well as to tails in the jet energy resolution.
Both D\O\ and CDF have presented new measurements of inclusive-jet 
rates using midpoint cone algorithms with a cone radius of $R=0.7$,
and based on approximately 1~fb$^{-1}$ of data.  

%%%%%%%%%%%%%%%%%%%%
%FIGURE
\begin{figure}
\centerline{\epsfxsize 3.4 truein \epsfbox{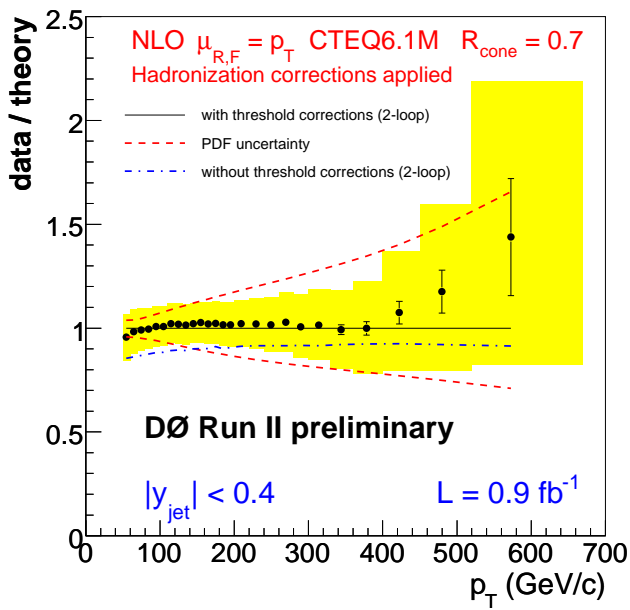}}
  \caption{Ratio of inclusive-jet data from D\O\
to theory for the midpoint cone algorithm ($R=0.7$)
and central jets, plotted versus transverse momentum $p_{\rm T}$.
Figure from ref.~\cite{D0InclJet07}.}
\label{D0incljetfigure}
\end{figure}

\begin{figure}
\centerline{\epsfxsize 3.6 truein \epsfbox{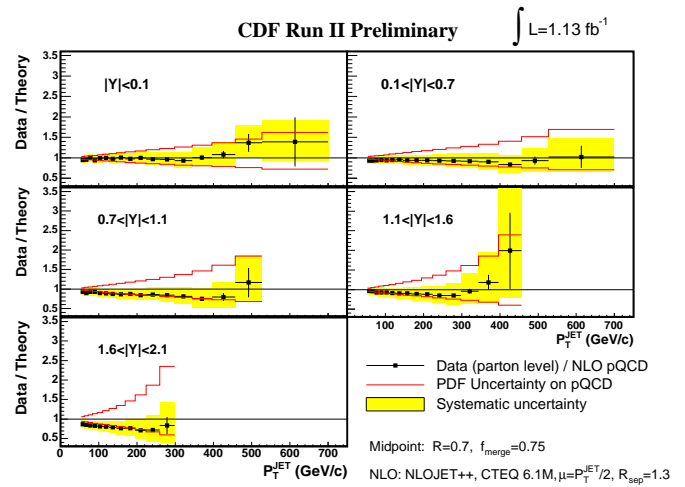}}
  \caption{Ratio of inclusive-jet data from CDF to NLO theory
for the midpoint cone algorithm ($R=0.7$), 
plotted versus $p_{\rm T}$ for various bins in rapidity $Y$.  
Figure from ref.~\cite{CDFInclJet07}.}
\label{CDFmidpointfigure}
\end{figure}

\Fig{D0incljetfigure} shows data from D\O\ divided by theory,
for central jets ($|y_{\rm jet}| < 0.4$) and binned in $p_{\rm T}$.
The theory includes NLO, plus an estimate of the NNLO terms based
on threshold logarithms computed at next-to-leading logarithmic (NLL)
accuracy~\cite{KOthr}.  NLO theory without these corrections
is about 10\% lower in this $p_{\rm T}$ range.  
The central value of the cross section from D\O\
is closer to the threshold-enhanced theory; however, 
the systematic uncertainty is too large to distinguish the 
two curves.
\Fig{CDFmidpointfigure} shows that the analogous data from 
CDF~\cite{CDFInclJet07} is also in good agreement with 
NLO theory~\cite{NLOJET}, at both central and forward
rapidities.

For both the D\O\ and CDF results,
the systematic uncertainty is dominated by the jet energy scale 
uncertainty and exceeds the statistical error for
essentially all points. However, the uncertainty arising from
the pdfs is even larger, and increases with $p_{\rm T}$.
This fact illustrates that the 
high-$p_{\rm T}$ jet data provide a strong constraint on pdfs, 
and can be used to reduce the pdf uncertainties in particular
regimes, particularly the gluon distribution $g(x)$ at large $x$.

%%%%%%%%%%%%%%%%%%%%
%FIGURE

\subsection{Di-jet azimuthal distributions}

At leading order in QCD, two jets produced in a $p\bar{p}$ collision
should emerge back-to-back in the plane transverse to the beam
axis, {\it i.e.} with the maximum azimuthal angle between them, 
$\Delta\phi = \pi$. QCD radiation pushes the azimuthal angle to smaller values.
In a multi-jet event, the azimuthal angle is defined to be between 
the two highest $p_{\rm T}$ jets.
For $\Delta\phi\approx\pi$, multiple soft gluon radiation dominates,
and fixed-order perturbation theory cannot be trusted.
Because of the way the azimuthal angle is defined, three parton
final-states can only produce $\Delta\phi > 2\pi/3$.
The $\Delta\phi$ distribution is an excellent test of how
well QCD describes complex hadronic final states, because
it is relatively insensitive to the overall jet energy scale
and to the pdfs.

\begin{figure}
\centerline{\epsfxsize 3.4 truein \epsfbox{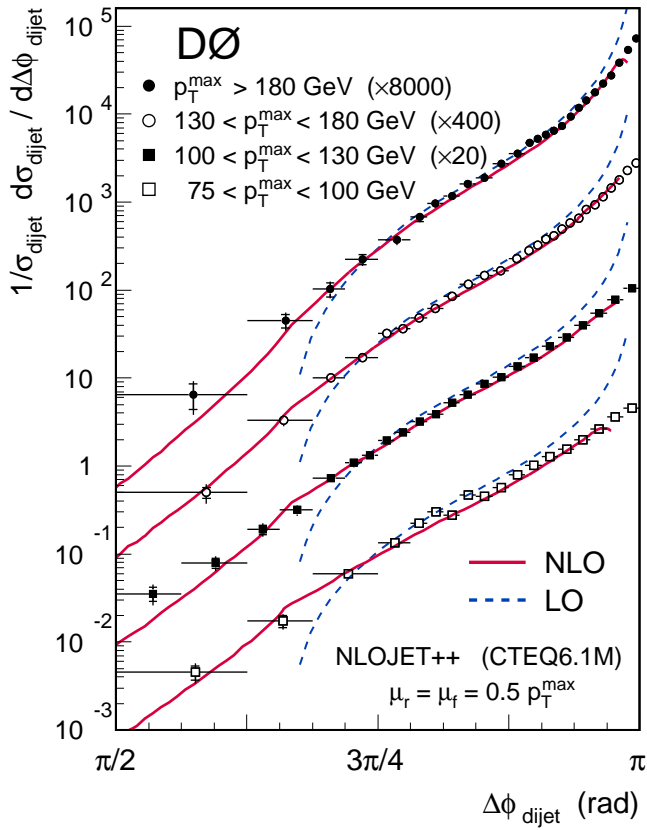}}
  \caption{Distribution in azimuthal angle $\Delta\phi$
between the two highest-$p_{\rm T}$ jets in an event,
measured by D\O\ for various $p_{\rm T}$ ranges,
and compared with LO and NLO QCD.  Figure from ref.~\cite{D0aziref}.}
\label{D0azifigure}
\end{figure}

\Fig{D0azifigure} shows D\O's measurement of the azimuthal decorrelation
in di-jet events at the Tevatron~\cite{D0aziref}, compared to LO and NLO
predictions from {\tt NLOJET++}~\cite{NLOJET}. 
LO theory fails for both small and large values of $\Delta\phi$,
for the reasons mentioned above.  However, NLO theory does very
well for $\Delta\phi < 2\pi/3$ (even though it is effectively
an LO calculation in this regime), and also pushes the agreement
for large values considerably closer to $\pi$.

%%%%%%%%%%%%%%%%%%%%
%FIGURE

\begin{figure}
\centerline{\epsfxsize 1.7 truein \epsfbox{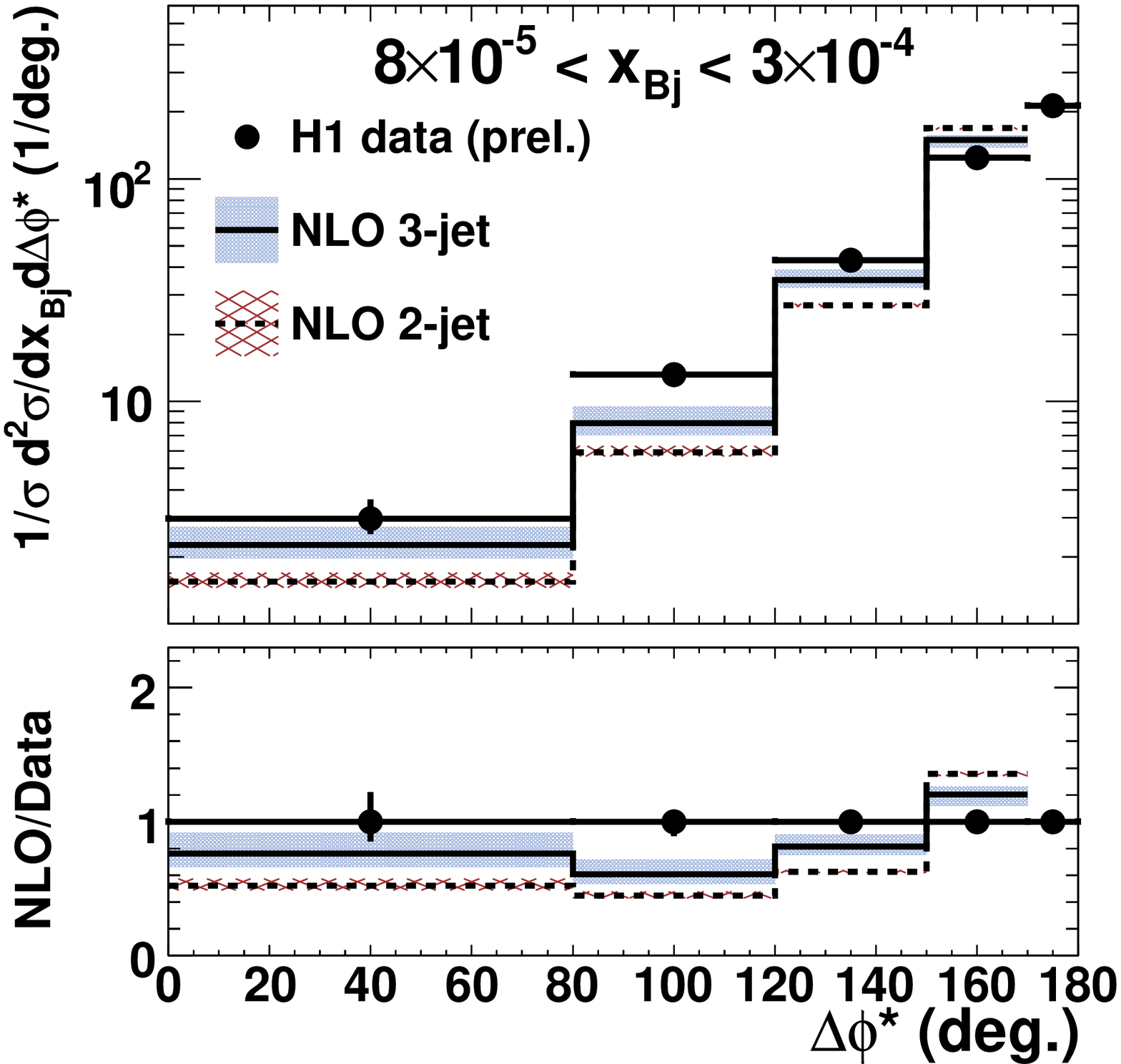}\ 
\epsfxsize 1.6 truein\epsfbox{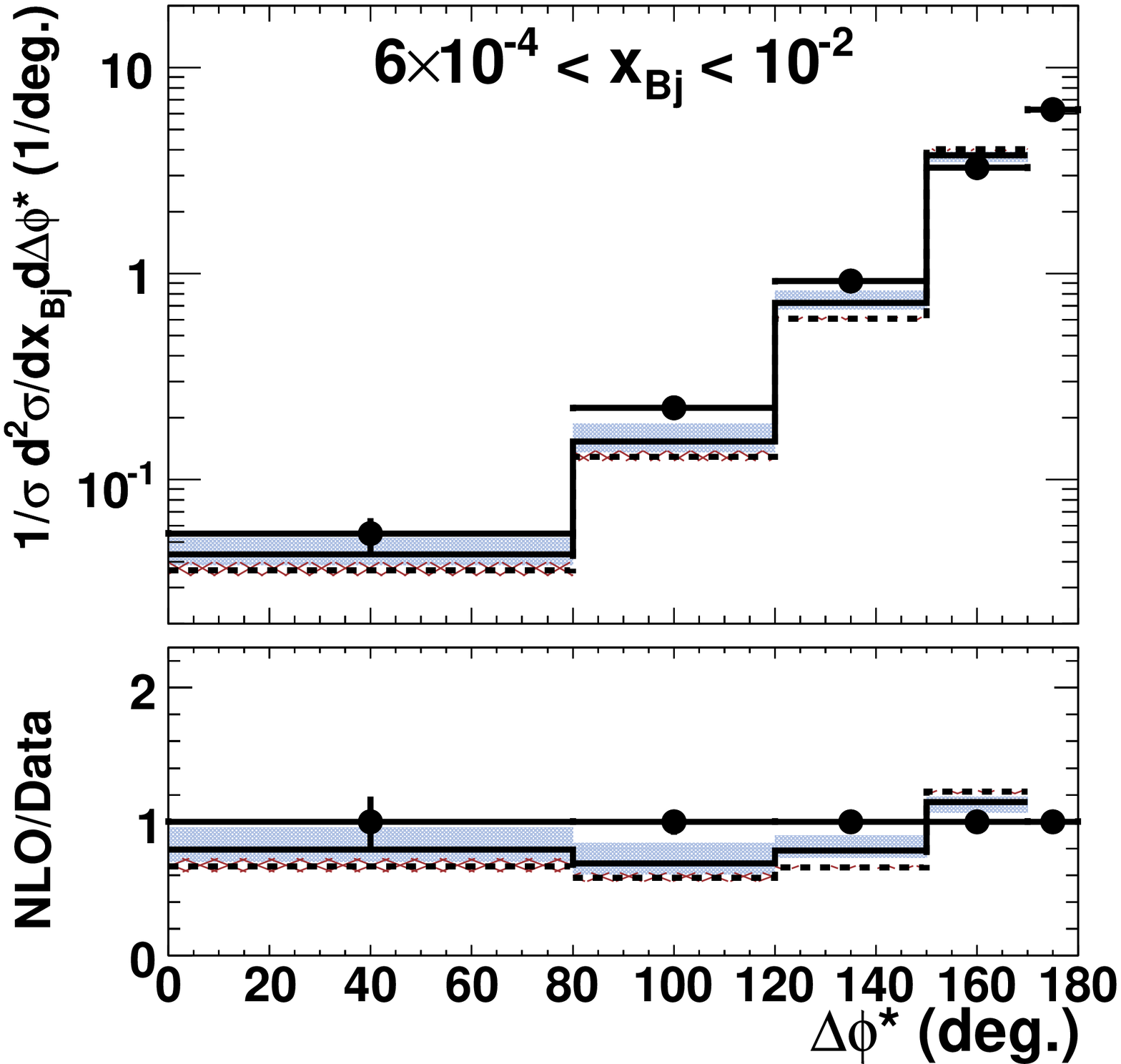}}
  \caption{Azimuthal decorrelations from H1 in the forward region.
Figure from ref.~\cite{H1aziref}.}
\label{H1azifigure}
\end{figure}

In contrast, NLO theory is unable to describe the azimuthal
correlations measured in $ep$ collisions by H1~\cite{H1aziref},
shown in \fig{H1azifigure}, at least for small values of $x_{\rm Bj}$.
The NLO three-jet predictions~\cite{epNLO3jet}
are closer to the data than are the two-jet predictions, 
but they still do not produce enough decorrelation at small 
$x_{\rm Bj}$.  Does this signal a breakdown of fixed-order
perturbation theory for these kinematics, and a need for
small-$x$ resummation?  Perhaps.
On the other hand, H1 also compared the data with 
a few different models incorporating small-$x$ resummations,
and all such models were found to be too low for small $\Delta\phi$
as well.

%%%%%%%%%%%%%%%%%%%%
%FIGURE

%%%%%%%%%%%%%%%%%%%%%%%%%%%%%%%%%

\section{Vector bosons plus jets}
\label{VSection}

\begin{figure}
\centerline{\epsfxsize 3.2 truein \epsfbox{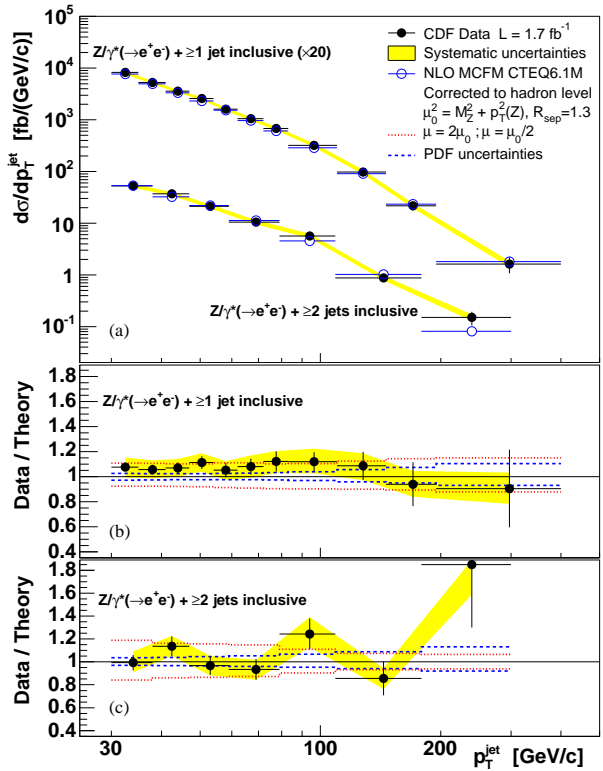}}
  \caption{Differential cross sections for $Z/\gamma^*(\to e^+e^-)$
+ 1 or 2 jets at the Tevatron, as a function of $p_{\rm T}^{\rm jet}$,
and in comparison with NLO theory~\cite{CampbellEllisVjj}. 
Figure from ref.~\cite{CDFZjj}.}
\label{cdfzjjfigure}
\end{figure}

The production of a single vector boson in association
with multiple jets is a background to searches for supersymmetry
at the LHC.  For example, the final state $Z+n$ jets, when the 
$Z$ decays to $\nu\bar\nu$, is a background to the 
missing-transverse-momentum plus multi-jet searches.
CDF and D\O\ have studied such events at the Tevatron,
for the decays $W\to\ell\nu$ and 
$Z\to\ell^+\ell^-$, plus up to 4 jets in the final state~\cite{Hesketh}.
Recently, CDF presented data~\cite{CDFZjj} on $Z$ plus 1, 2 and 3 jets,
in comparison with fixed-order LO and NLO predictions.
\Fig{cdfzjjfigure} shows the excellent absolute agreement, to within 10\%, 
with NLO predictions for 1 and 2 jets, the maximum number for which 
NLO results are currently available~\cite{CampbellEllisVjj}.
These results highlight the importance of extending NLO theory to 
larger numbers of jets, in order to be ready for the enormous 
data sets that will be available in these channels at the LHC.

%%%%%%%%%%%%%%%%%%%%
%FIGURE

%%%%%%%%%%%%%%%%%%%%%%%%%%%%%%%%%

\section{Top quarks plus jets}
\label{TSection}

The final-state $t\bar{t}+$jet is another important background
to supersymmetry at the LHC.  The cross section is large,
and the additional jet can boost the $t\bar{t}$ system so that
neutrinos from top quark decay generate large missing transverse
momentum.  The NLO corrections to this process were computed
recently~\cite{ttDUW}.  They require the evaluation of many 
virtual Feynman diagrams, including pentagon diagrams with massive
propagators, and a large number of subtraction terms for the real
corrections~\cite{CDST}.  The results greatly reduce the uncertainty
on the $t\bar{t}+$jet cross section at the LHC.

At the Tevatron, a $p\bar{p}$ collider, the forward-backward 
asymmetry of $t\bar{t}$ pairs is an interesting observable that
probes the dynamics of top quark production.  It can be defined by
\be
A_{\rm FB}^{t} \ =\ { N_t(y_t>0) - N_t(y_t<0)
                 \over N_t(y_t>0) + N_t(y_t<0) }\,,
\label{AFBtDef}
\ee
where $N_t(y_t>0)$ is the number of top quarks produced
with positive rapidity. This quantity vanishes at leading order, 
$\Ord(\alpha_s^2)$, just like the forward-backward asymmetry 
for muons in the QED process $e^+e^-\to\mu^+\mu^-$.
But it is nonvanishing at NLO, or 
$\Ord(\alpha_s^3)$~\cite{ttFBearly,ttBER,AKR07},
as shown in table~\ref{ttFB}.
It is also nonvanishing for the process $t\bar{t}+$jet, computed 
at its leading order, again $\Ord(\alpha_s^3)$. 
Here a jet is required to have $p_{\rm T}^{\rm jet} > 20$~GeV.
Ref.~\cite{ttDUW} computed the asymmetry~(\ref{AFBtDef})
in $t\bar{t}+$jet events at NLO, $\Ord(\alpha_s^4)$,
and found the striking result that the asymmetry is drastically
reduced, from $-6.9\%$ to essentially zero.  One might wonder 
what this result portends for the NNLO value of the $t\bar{t}$
inclusive asymmetry --- will it too receive a large correction?

%%%%%%%%%%%%%%%%%%%%%%%%%%%%%%%%%%%%%%
\begin{table}
\caption{\label{ttFB} $t\bar{t}$ forward-backward asymmetry 
  at the Tevatron, from refs.~\cite{ttBER,ttDUW}.}

\vskip .4 cm

\begin{tabular}{||c||c|c|c||}
\hline
\hline
$A_{\rm FB}^t(\%)$ & $t\bar{t}$ inclusive
         & $t\bar{t}$+jet inclusive & $t\bar{t}$0j exclusive \\
% five
\hline
\hline
LO  & 0 & $-$6.9 & 0 \\
\hline
NLO & 3.8 & $-1.5\pm1.5$ & 6.4\\
\hline
\hline
\end{tabular}
%\end{ruledtabular}
\end{table}

In the mean time, the first forward-backward asymmetry measurements
from the Tevatron were reported at this conference.
The observable used by CDF~\cite{ErbacherLP07} 
and D\O~\cite{ttD0,ErbacherLP07} is a bit different from
\eqn{AFBtDef}. It employs the rapidity difference between
the $t$ and $\bar{t}$,
\be  
A^{t} \ =\ { N_t(y_t>y_{\bar{t}}) - N_t(y_t<y_{\bar{t}})
                 \over N_t(y_t>y_{\bar{t}}) + N_t(y_t<y_{\bar{t}}) }\,.
\label{AtDef}
\ee
The difference between these two observables for the NLO inclusive
definition was studied recently~\cite{AKR07}.
For the same choice of pdfs, cuts, {\it etc.}, it is found that
$A_{\rm FB}^{t}=(5.1\pm0.6)\%$, whereas $A^{t}=(7.8\pm0.9)\%$.
The observable $A^{t}$ is somewhat larger than $A_{\rm FB}^{t}$ because
events where both the $t$ and $\bar{t}$ go forward can 
contribute to $A^{t}$ but not to $A_{\rm FB}^{t}$.

CDF measures, with 1.7~fb$^{-1}$ of data,
\be
A^{t} = (28\pm13_{\rm stat}\pm5_{\rm syst})\%\,; 
\label{CDFAt}
\ee
while D\O\ measures, with 0.9~fb$^{-1}$ of data,
\be
A^{t}_{\rm uncorr} = (12\pm8_{\rm stat}\pm1_{\rm syst})\%\,,
\label{D0At}
\ee
the latter number has not been corrected for 
reconstruction effects.  The statistical errors are still 
large, of course.  However, in light of the large central values,
it will certainly be interesting to follow these
results as more data are analyzed.

%%%%%%%%%%%%%%%%%%%%%%%%%%%%%%%%%
\section{Conclusions}
\label{ConclusionsSection}

This talk has surveyed some of the recent progress over the past
year or so in our quantitative theoretical understanding of 
hard QCD processes at colliders.  An increasing number of processes
are now known at NLO, and for a few benchmark processes, NNLO 
precision is available.   Experiments at the Tevatron and at HERA
test such QCD predictions over a wide range of kinematics.
The experiment between experiment and theory is generally very good,
except near kinematic boundaries such as small transverse momentum 
and small $x$, for which resummations and reorganizations of the
perturbation theory should be performed.  
It is critical to push the ``loops and legs'' frontier to bring
more processes to NLO and NNLO accuracy, and also to incorporate
such processes into parton showers, while retaining that accuracy.
But in general, we are ``virtually'' ready for the startup of 
the LHC next year! 

%%%%%%%%%%%%%%%%%%%%%%%%%%%%%%%%%%%%%%%%%%%%%%%%
%% BACKMATTER
%%%%%%%%%%%%%%%%%%%%%%%%%%%%%%%%%%%%%%%%%%%%%%%%

\vskip0.3cm
\par\noindent
{\bf Acknowledgments}

I am grateful to Z. Bern, D. Brown, M. Diehl, R. Erbacher, C. Gwenlan, 
J. Huston, M. Peskin, G. Salam and K. Tollefson for assistance in 
preparing this talk and write-up.  I also thank the organizers of 
Lepton-Photon 2007 for the opportunity to present the talk
at such a stimulating conference.

%%%%%%%%%%%%%%%%%%%%%%%%%%%%%%%%%%

\end{document}